\documentclass[]{spie}  

\usepackage{amsmath,amsfonts,amssymb}
\usepackage{graphicx}
\usepackage[colorlinks=true, allcolors=blue]{hyperref}

\title{A New Method for Wavefront Sensing using Optical Masking Interferometry}

\author[a]{Christopher L. Carilli}
\author[b]{Laura Torino}
\author[c]{Bojan Nikolic}
\author[d]{Nithyanandan Thyagarajan}
\author[b]{Ubaldo Iriso}

\affil[a]{National Radio Astronomy Observatory, Socorro, NM, USA, 87801}
\affil[b]{ALBA - CELLS Synchrotron Radiation Facility\\Carrer de la Llum 2-26, 08290 Cerdanyola del Vallès (Barcelona), Spain}
\affil[c]{Astrophysics Group, Cavendish Laboratory, Univ. of Cambridge, Cambridge CB3 0HE, UK}
\affil[d]{Commonwealth Scientific and Industrial Research Organisation (CSIRO), Space \& Astronomy, P. O. Box 1130, Bentley, WA 6102, Australia}

\authorinfo{Further author information: (Send correspondence to
Chris Carilli)\\E-mail: ccarilli@nrao.edu, Telephone: 1 575 835 7306}

\begin{document} 
\maketitle

\begin{abstract}
Wave front sensing of the surface of equal phase for a propagating electromagnetic wave is a vital technology in fields ranging from real time adaptive optics, to high accuracy metrology, to medical optometry. We have developed a new method of wavefront sensing that makes a direct measurement of the electromagnetic phase distribution, or path-length delay, across an optical wavefront. The method is based on techniques developed in radio astronomical interferometric imaging. The method employs optical interferometry using a 2-D aperture mask, a Fourier transform of the interferogram to derive interferometric visibilities, and self-calibration of the complex visibilities to derive the voltage amplitude and phase gains at each hole in the mask, corresponding to corrections for non-uniform illumination and wavefront distortions across the aperture, respectively. The derived self-calibration gain phases are linearly proportional to the electromagnetic path-length distribution to each hole in the aperture mask, relative to the path-length to the reference hole, and hence represent a wavefront sensor with a precision of a small fraction of a wavelength. The method was tested at $\lambda=400\,$nm at the Xanadu optical bench at the ALBA synchrotron light source using a rotating mirror to insert tip-tilt changes in the wavefront. We reproduce the wavefront tilts to within $0.1''$ ($5\times 10^{-7}$~radians). We also derive the static metrology though the optical system for non-planar wavefront distortions to $\sim \pm1$~nm repeatability. Lastly, we derive frame-to-frame variations of the wavefront tilt due to vibrations of the optical components which range up to $\sim 0.5"$. These variations are relevant to adaptive optics applications. Based on the measured visibility phase noise after self-calibration, we estimate an rms path-length precision per 1~ms exposure of 0.6 nm. 
\end{abstract}


\section{INTRODUCTION}
\label{sec:intro}  

Wave front sensing (WFS) of the surface of equal phase for a propagating electromagnetic wave\cite{geary1995} is a vital technology in fields ranging from real time adaptive optics in laser communications and directed energy systems\cite{Watnik18,holmes2022}, to high precision metrology and manufacturing\cite{forest2004, rammage2002, adapa2020}, to medical ophthalmology\cite{Liang1994, neal2002}. In this paper, we present a new method of optical wavefront sensing of the full shape of the wavefront using interferometric techniques developed in radio astronomy. 

Our technique employs interferometric imaging with a 2-D aperture mask, plus Fourier inversion of the interferogram to derive the spatial mutual coherences per interferometric baseline, or visibilities, in the aperture plane. We employ a non-redundantly sampled mask to mitigate decoherence \cite{Readhead+1988}. We then apply self-calibration techniques from radio astronomy for a joint derivation of the source structure and the complex voltage gains at each interferometric element in the aperture plane, i.e. hole in the mask \cite{Schwab1980, Schwab1981, Readhead+Wilkinson1978, Cornwell+Wilkinson1981}. The focus of our previous work was on deriving the 2D structure of the electron beam at the ALBA synchrotron light source. We relied on the known Gaussian nature of the synchrotron source to limit the self-calibration to just voltage amplitudes, from which we derive the illumination pattern across the mask and the source Gaussian parameters, both to better than 1\% precision\cite{Nikolic2024, iriso2024, Thyagarajan+2025}. 

In this report, we expand the Fourier analysis to include self-calibration of the full complex visibilities (real and imaginary or amplitude and phase), which provides both a measurement of the source size and shape, as well as of the element-based voltage gain amplitude and phase corrections. The key point for this report is that the element-based gain phases provide a linear measure of the photon path-lengths through the optical system to the aperture plane relative to a reference position in the mask, and hence represent an accurate wavefront sensor for the distortions across the wavefront. Including both phase and amplitude gain solutions allows for a more general approach to the Fourier imaging, self-calibration, and source modeling process. Such a more general analysis has been applied to sources of arbitrarily complex morphological shape in radio astronomy \cite{Cornwell+Fomalont1999, Pearson+Readhead1984, Readhead+Wilkinson1978, Schwab1981, Perley1999, TMS2017, Cornwell+Fomalont1999, perley1984}. 

Herein, we present tests of the method using a rotating mirror in the optical path at ALBA. The rotating mirror allows for introduction of a tip-tilt distortion to the wavefront, which can then be measured using the self-calibration technique. We also derive the non-planar static path-length, or metrology, terms for the wavefront, as well as the frame-to-frame wavefront variations relevant to adaptive optics. 

We focus initially on the dominant low order mode of the wavefront, namely, the tip-tilt, because the interferometric mask employed is still relatively sparse (7-holes), and because there are two independent methods to derive the wavefront tilt using the same data which provide a critical calibration of the results. 

\section{Wavefront Sensing}
\label{sec:WFS}

We briefly summarize some basic aspects of wavefront sensing. An electromagnetic wavefront is defined as the surface perpendicular to the direction of propagation of the EM wave, corresponding to the surface of equal phase. Wavefront distortions due to non-ideal optical components, or due to turbulence in the laboratory atmosphere, can lead to `corrugations' in an initially planar wavefront, corresponding to path-length delays across the wavefront as it transverses the optical system (see Figure~\ref{fig:OpticsWFS}). Wavefront sensing corresponds to methods designed to determine these path-length delays \cite{geary1995}.  

Wavefront sensing has a wide range of uses, including: (i) real-time adaptive optics to correct for turbulence or vibrations of optical components in astronomical, laboratory, and industrial systems \cite{davies2012, Watnik18, holmes2022}, (ii) surface metrology of physical components to high accuracy \cite{forest2004, rammage2002, adapa2020}, and (iii) medical optometry to determine the shape of the cornea and other optical path elements in the eye \cite{rammage2002, neal2002, Liang1994}. There are numerous methods in use today for wavefront sensing, threee of the most prominent being: (i) reference  wave interference methods (Fizeau\cite{sheldovka,goodwinwyant}), (ii) angle-based, or gradient-based wavefront sensors (Shack-Hartmann\cite{neal2002,Mansuripur2009,yi2021}), and (iii) generalized Foucault knife edge tests (pyramid\cite{shatokhina}). 

The reference wave technique is based on Fizeau-interferometry, where beam splitting is used to interfere the target wavefront with a 'reference wavefront', either a copy of the original wavefront reflected off a known reference surface, or an independently generated reference wavefront. This method requires a well calibrated and stable reference surface, or a second, well understood light source, in order to determine the target wavefront distortions. 

The Shack-Hartmann wavefront sensor involves a lenslet aperture array which images a planar incoming wavefront into a uniform grid of sources in the focal plane. Wavefront corrugations distort this grid due to the different angle of incidence at each lenslet. The local offsets of a source from the expected grid point determine the phase gradient of the incident plane wave across that particular lenslet. The Shack-Hartmann sensor therefor measures the wavefront gradient distribution across the wavefront i.e. the first derivative. Fitting of smooth functions to the gradient distribution, such as 2D Zernike polynomials, is then used to determined the dominant modes for path-length distortions of the wavefront across the aperture \cite{vacalebre2022}. Shack-Hartmann sensors have been developed with nanometer accuracy \cite{adapa2020, neal2002}, and with dynamic ranges in the hundreds, or even a thousand, where dynamic range corresponds to the largest optical path-length distortion measurable, in units of wavelengths \cite{Liang1994, Lombardo2009}. 

The pyramid wavefront sensor is essentially a Foucault knife edge test generalized to four simultaneous knife edges using a moving pyramid prism near the focus\cite{shatokhina}. These are widely used in astronomical applications. 

In the following section, we show that the self-calibration phase solutions in our processing represent a new approach to wavefront sensing, in which the gain phases, $\phi_G$, are linearly proportional to wavefront path-length delays across the mask, $\rm \delta L(t)$, relative to the reference aperture element, or hole in the mask. Hence, the shape of the wavefront is measured naturally through phase self-calibration. Our method measures the full wavefront across the aperture (i.e. not a gradient method), at least at the sampled positions of the mask. Our method does not require a reference beam (Fizeau), nor does it have any moving parts (pyramid). 

\section{Basics of Interferometric (Fourier) Imaging and Self-Calibration}
\label{sec:method}

Interferometric imaging and self-calibration techniques have been well documented in the context of radio astronomical measurements\cite{TMS2017}. We summarize the basics for completeness.

The spatial coherence (or \textit{visibility}), $V_{ab}(\nu)$,
corresponds to the cross correlation of two quasi-monochromatic voltages of frequency $\nu$ of the same polarization sensed by two spatially distinct elements in the aperture plane of an interferometer. The visibility relates to the intensity distribution of an incoherent source in the far field, $I(\hat{\mathbf{s}},\nu)$, via the van Cittert-Zernike theorem  \cite{vanCittert34, Zernike38}, which after certain approximations reduces to a Fourier transform \cite{TMS2017,Born+Wolf1999}:

\begin{align}
    V_{ab}(\nu) &= \int_\textrm{source} A_{ab}(\hat{\mathbf{s}},\nu) I(\hat{\mathbf{s}},\nu) e^{-i2\pi \mathbf{u}_{ab}\cdot \hat{\mathbf{s}}} \mathrm{d}\Omega \, , \label{eqn:VCZ-theorem}
\end{align}

\noindent where, $a$ and $b$ denote a pair of array elements (eg. holes in a mask), $\hat{\boldsymbol{s}}$ denotes a unit vector in the direction of any location in the image, $A_{ab}(\hat{\mathbf{s}},\nu)$ is the spatial response (the `power pattern') of each element (in the case of circular holes in the mask, the power pattern is the Airy disk, with a size inversely proportional the hole diameter), $\mathbf{u}_{ab}=\mathbf{x}_{ab} (\nu/c)$ is the ``baseline'' vector = the vector spacing ($\mathbf{x}_{ab}$) between the element pair in units of wavelength (or inverse radians), and $\mathrm{d}\Omega$ is the differential solid angle element on the image (focal) plane. 

In practice, a visibility is a complex-valued measurement in the aperture plane defined by $u$--$v$ coordinates, which, after Fourier transform, implies a sinusoidal (`fringe') pattern in the image-plane, with a spatial frequency and orientation determined by the baseline vector, and having an amplitude corresponding to the power of the source mutual coherence at that fringe spacing, and a phase corresponding to the position of that fringe relative to the adopted phase center. This concept of a sinusoidal fringe in an image from an interferometric baseline has been the basis of interferometry since Young's original 2-slit experiment of 1803\cite{young1803}.

We employ optical aperture masking interferometry, in which a mask with a specific pattern of holes, or interferometric elements, is placed in the aperture plane. Our current experiment employs a 7-hole non-redundant aperture mask based on a scaled and stretched version of the aperture mask used on the James Webb Space Telescope \cite{JWST}. Non-redundancy means that each baseline in the mask (vector hole separations), is unique, thereby avoiding decoherence of redundant visibilities due to phase fluctuations across the aperture. Light passing through the aperture mask is focused by a lens through reimaging optics, a polarizer, and a narrow band filter, and onto a CCD which records the interferogram. The complex visibilities are generated via a Fourier transform of the interferogram. 

\begin{figure}[!htb]
\centering 
\centerline{\includegraphics[scale=0.25]{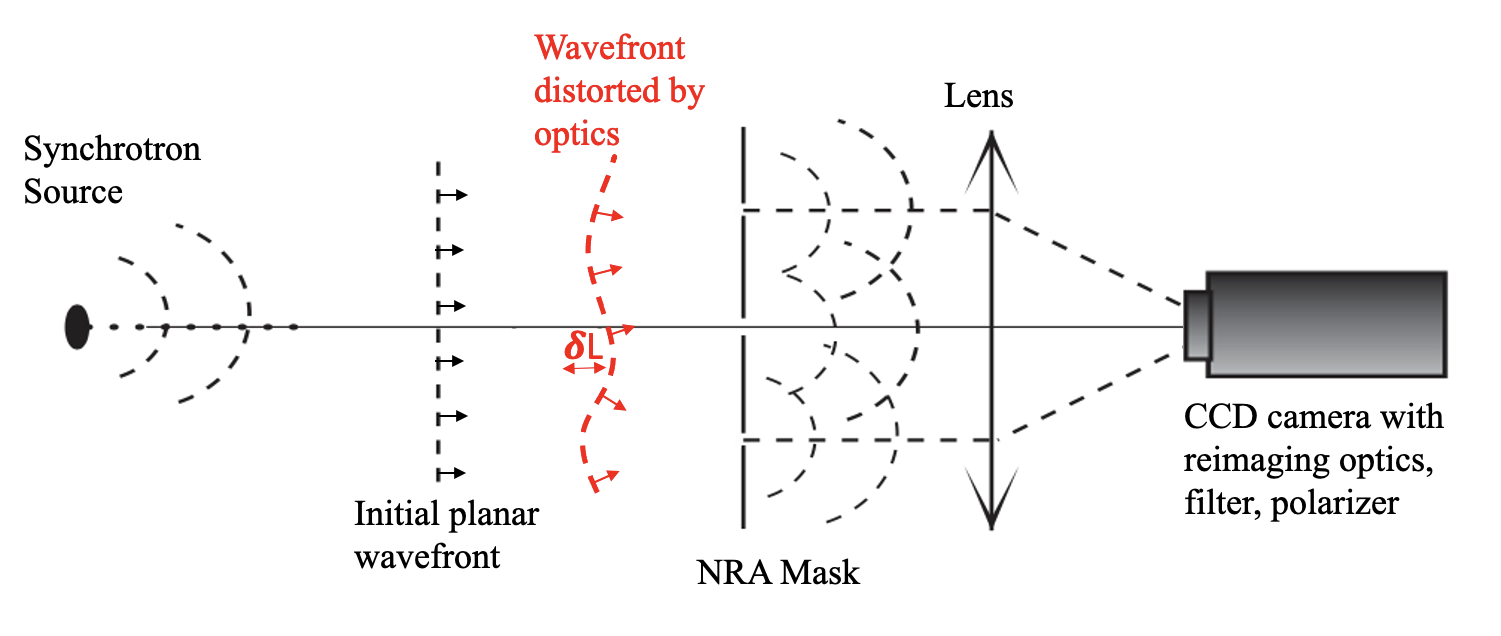}}
\caption{Schematic of the optical system with cartoon of wavefront distortions. Note that multiple mirrors in vacuum and out are not shown, including the rotating mirror (see Figure~\ref{fig:rotmir}). The rotating mirror can be considered a planar 'distortion' of just the tip-tilt term for the wavefront. 
}
\label{fig:OpticsWFS}
\end{figure}

The measured visibilities can be corrupted by distortions introduced by the propagation medium, or the relative illumination of the holes, or other effects in the optics (vibrating or non-perfect mirrors or lenses). These distortions can be factorized into multiplicative element-based complex voltage gain factors, $G_a(\nu)$ and $G_b(\nu)$, where $a$ and $b$ represent the two interferometer elements in the visibility baseline. Thus, the corrupted measurements, $V_{ab}^\prime(\nu)$, are given by the complex product:
\begin{align}
    V_{ab}^\prime(\nu) &= G_a(\nu) \, V_{ab}(\nu) \, G_b^\star(\nu) \, , \label{eqn:uncal-vis}
\end{align}
where, $\star$ denotes a complex conjugation, and $V_{ab}(\nu)$ is the true source visibility. 

\begin{figure}[!htb]
\centering 
\centerline{\includegraphics[scale=0.28]{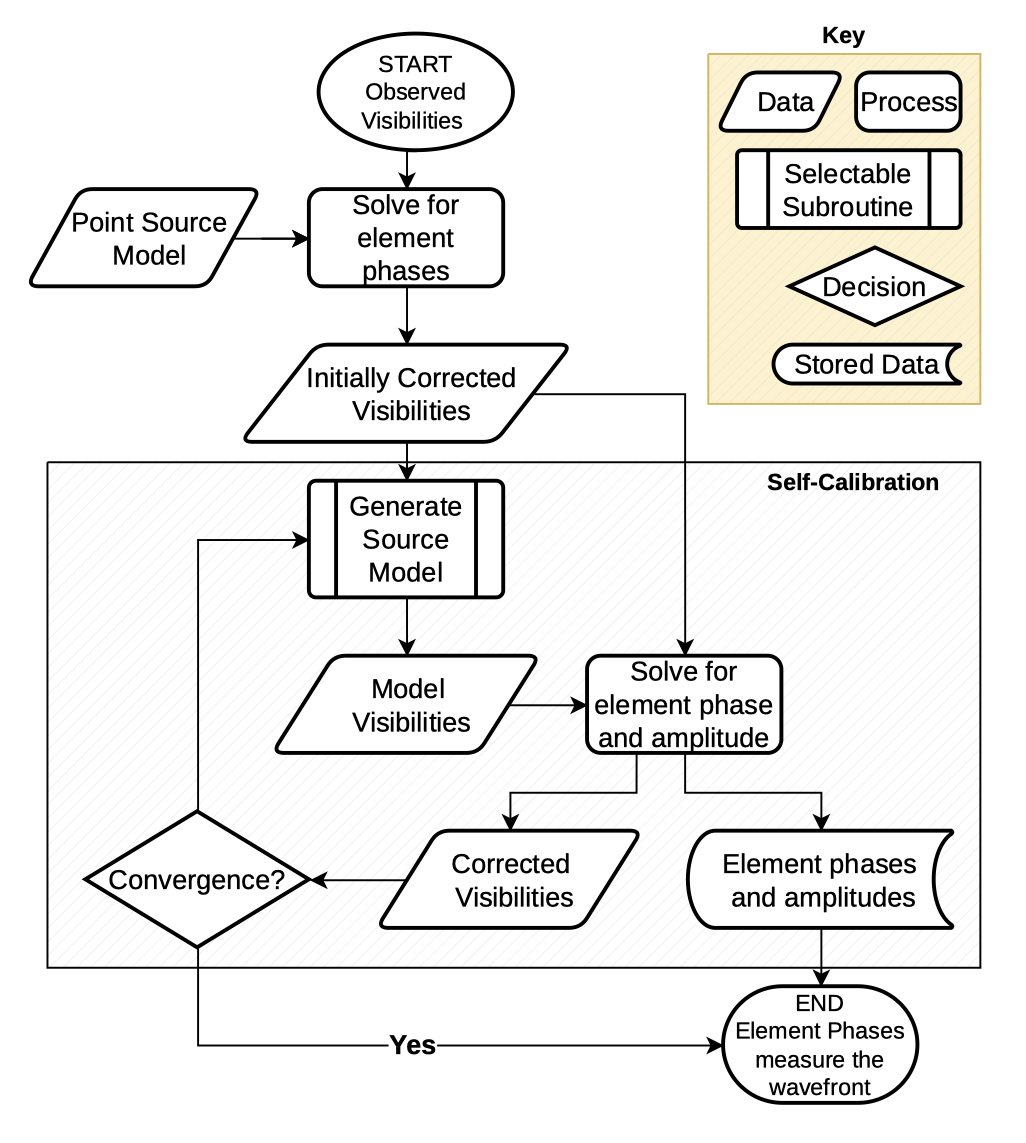}}
\caption{The self-calibration algorithm. The ``Generate Source Model'' subroutine can involve Gaussian fitting to the corrected visibilities described in Section~\ref{sec:method}, or other model fitting or deconvolution algorithms.}
\label{fig:selfcal}
\end{figure}

The process of interferometric self-calibration determines these complex voltage gain factors in parallel with determining source structure. A physically reasonable starting model for the source is assumed, $V_{ab}(\nu)$. Using the measurements, $V_{ab}^\prime(\nu)$, Equation~(\ref{eqn:uncal-vis}) is then inverted to derive the complex voltage gains, $G_a(\nu)$, using an optimization criterion, such as least squares fitting \cite{Schwab1980, Schwab1981, Readhead+Wilkinson1978, Cornwell+Wilkinson1981, Nikolic2024}. A new source model is then derived from the gain-corrected visibilities through model fitting or imaging (Fourier inversion) and deconvolution, and the process is iterated until convergence. A block diagram of the self-calibration process is shown in Figure~\ref{fig:selfcal}.

The self-calibration process converges since there are typically more measurements ($[N(N-1)]/2$ complex cross correlations, or visibilities, where $N$ is the number of interferometric elements), than fitted parameters ($N$ element-based complex gains, and model parameters), at least for reasonably sampled visibility data and a not-too-complex source \cite{Schwab1980, Schwab1981}. 

Our previous papers using this system have focused on derivation of the source size and shape, in which the self-calibration is a critical step for correcting illumination distortions, but the gain solutions are not of immediate interest \cite{Nikolic2024, iriso2024, Thyagarajan+2025}. In this paper, focus is on the self-calibration solutions themselves. In particular, the voltage hole-based phase solutions, or `phase gains',  $\phi_G$, represent a new approach to wavefront sensing, in which $\phi_G$ is linearly proportional to wavefront path-length delays to the aperture, $\rm \delta L(t)$, relative to the reference hole\footnote{Phase, or path-length delay, is always a difference measurement with respect to a reference position since the wavefront is comprised of millions of incoherent photons, for which absolute phase is ill-defined, but phase difference between spatial positions in the mask remains a well defined and physical quantity common among photons, thereby allowing for mutual coherence and application of the van Cittert-Zernike theorem.}, through the simple relation: 

\begin{align}
\rm \delta L = \lambda \times (\phi_G /360^\circ), \label{eqn:path}
\end{align}

\noindent where $\rm \phi_G$ is in degrees. Hence, the gain phases correspond to an accurate (small fraction of a wavelength), high time resolution wavefront sensor.  This concept is shown schematically in Figure~\ref{fig:OpticsWFS}. 

\section{Experiment}
\label{sec:experiment}

The details of the experimental set-up are given in previous papers in this series\cite{Nikolic2024, iriso2024}. We briefly summarize the set-up and processing for completeness. Figure~\ref{fig:OpticsWFS} shows a simplified schematic of the synchrotron radiation interferometry (SRI) system optics, without showing the full optical path, which includes 8 flat mirrors before the mask \cite{Torino2016}. The total path-length from the mask to the synchrotron source is 15.0~m. 
The synchrotron optical source is extraordinarily bright, with of order $10^7$ photon counts in each interferogram per 1~ms exposure. For reference, the Gaussian parameters for the synchrotron light source are: major axis FWHM = $140~\mu$m ($1.9''$ at 15~m distance); minor axis FWHM = $56~\mu$m ($0.77''$); and major axis position angle of $24^\circ$\cite{Nikolic2024}.

\begin{figure}[!htb]
\centering 
\centerline{\includegraphics[scale=0.32]{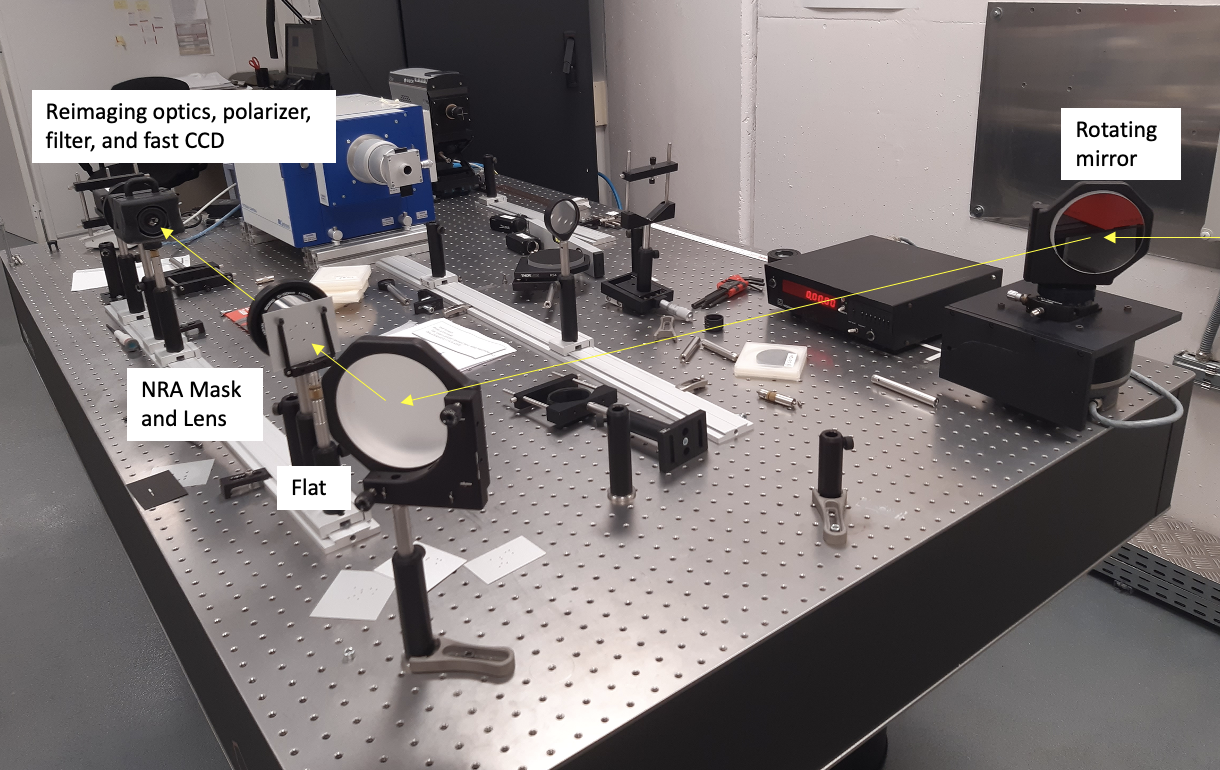}}
\caption{Picture of the optical bench setup at Xanadu, ALBA, showing the location of the rotating mirror. The photons from the synchrotron beam line enter from the right of the picture, as indicated by the yellow lines and arrows. NRA = non-redundant aperture mask. 
}
\label{fig:rotmir}
\end{figure}

Figure~\ref{fig:rotmir} shows the optical bench at Xanadu. The only difference between the current experiment and previous experiments is the insertion of a flat mirror that can be rotated by set increments. This rotating mirror allows us to introduce wavefront tip-tilt changes which can then be measured using the various techniques described in Section~\ref{sec:tilt}. A 7-hole non-redundant aperture mask with 2~mm diameter holes is employed. The maximum baseline (hole separation) on the mask is 22~mm. A CCD camera of 1296x966 pixels was employed to obtain interferograms of the diffraction pattern with a narrow band filter at 400~nm, with a pixel scale $0.153''$. Each frame exposure was 1~ms in duration, and each experiment entailed 30 frames taken with a frame separation of 1~s. 

A series of experiments were performed in which the rotating mirror was nominally adjusted as follows:  $0''$, $0''$, $1''$, $2''$, $3''$, $4''$, $0''$. The rotation is only done around the Y axis (vertical), meaning we expect wavefront tilts in the X direction (horizontal). The $0''$ experiments were repeated at the beginning, and then again at the end, to check hysteresis and repeatability. Note that on reflection, the wavefront tilt induced by the rotated mirror is twice the mirror rotation angle, so a $1''$ rotation of the mirror will induce a $2''$ wavefront tilt.

We emphasize that these rotation angles are small, eg. $1'' = 5\times 10^{-6}$~radians.  A nominal rough estimate for the accuracy of setting of our rotating mirror device is $\sim 0.3''$, which is inadequate to reach the level of testing we desire. In order to calibrate the measurements to higher accuracy, we have developed three independent methods to derive the wavefront tilt from the data: one in the image plane, a second using the measured visibilities, and the third being self-calibration. We find these three methods provide the critical cross-check, or calibration, of wavefront measurement to $\sim 0.1''$ precision (Section~\ref{sec:tilt}).

\section{Processing}
\label{sec:processing}

\begin{figure}[!htb]
\centering 
\centerline{\includegraphics[scale=0.25]{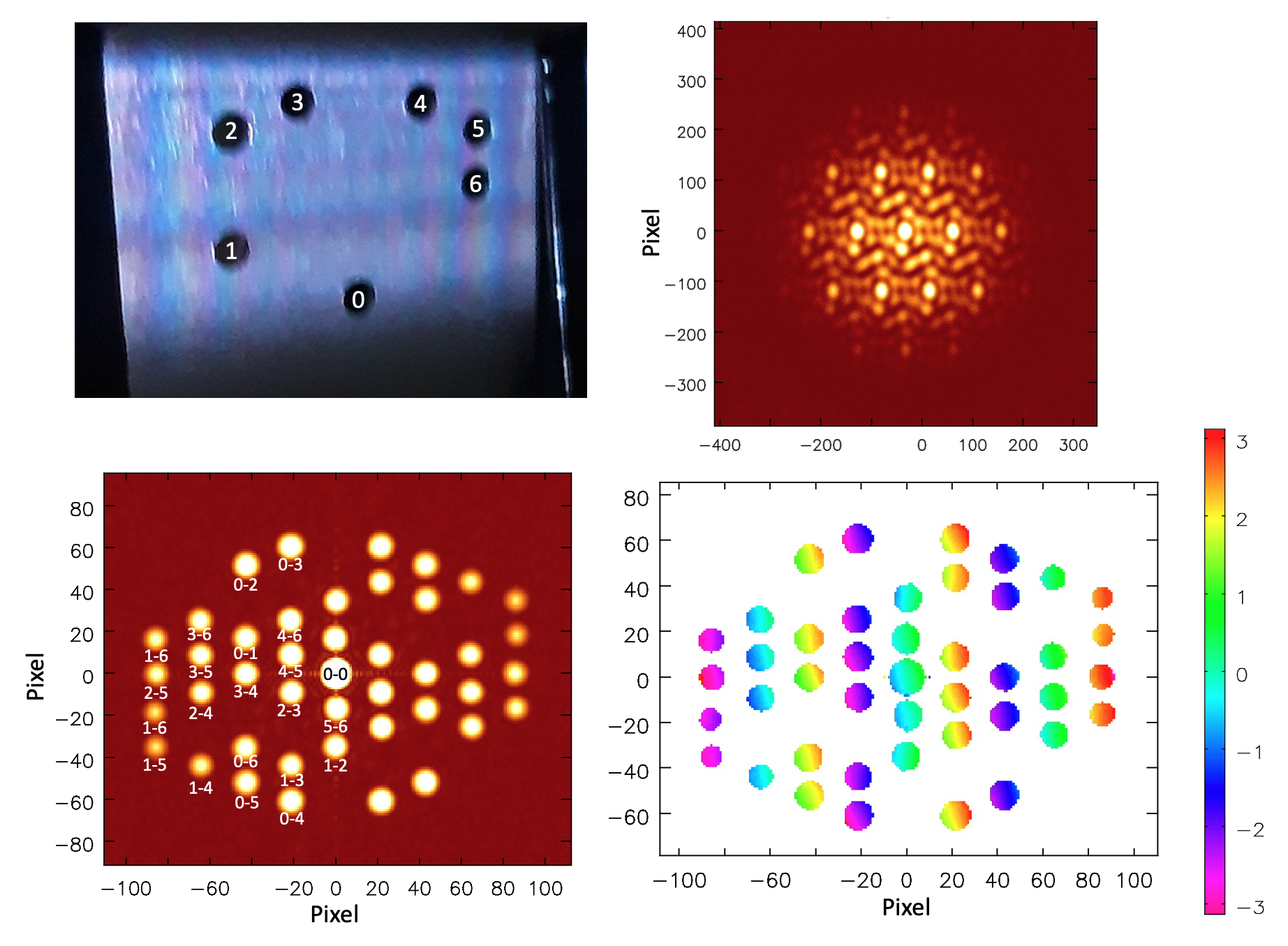}}
\caption{Top Left: Photograph of the 7 hole mask illuminated by the synchrotron light beam. Holes are 2~mm in diameter, and they are numbered. Top right: interferogram made using the 7 hole mask. The pixel scale is $0.153''$ per pixel, which, using the 15.03~m distance from the list source to the mask, corresponds to 11.1~$\mu$m in the source plane. Bottom left: Fourier transform of the interferogram showing the visibility amplitudes. Interferometric baselines between holes are numbered below each visibility sample (0-0 corresponds to the autocorrelation = total power; Hermitian conjugate samples are not numbered). The pixel scale is baseline length measured in wavelengths (or inverse radians), with 657 wavelengths per pixel. Bottom right: Same, showing the visibility phases. The color scale is in radians. Note that the sizes of the discrete regions in the $u$--$v$ coordinates is linearly related to the size of the holes, or the inverse size of the Airy disk\cite{Carilli2024}. 
}
\vspace*{0.5cm}
\label{fig:4frame}
\end{figure}

Figure~\ref{fig:4frame} shows the illuminated mask, and the image products from the various processing steps, including an interferogram and the resulting Fourier transform products. 

We employ the same Fourier processing steps to go from CCD images to visibilities as in previous work\cite{Nikolic2024}. In brief, images were bias corrected, and padded to 2048 pixels prior to Fourier transforming. Before transforming, each image was shifted to a nominal center. For the $0''$ rotation data, the centering process involved smoothing the interferogram on a scale of the Airy disk, or by about 50 pixels. The centering entails shifting the unsmoothed padded image to the peak position measured on the smoothed image. This Airy disk centering takes out the dominant tip-tilt term through the full optical system. Further, the centering pixel position can be used as a measurement of the wavefront tilt, as discussed in Section~\ref{sec:tilt}. We find that the Airy disk centering pixel repeats to 2 pixels or $0.3''$ for the three 0" rotation data (Section~\ref{sec:experiment}).

For the experiments in which the mirror is rotated, we employ the mean centering pixel derived for the three unrotated experiments, ie. $0''$. In this way, we remove the overall system tip-tilt, but retain the tilt imposed by the rotating mirror. 

Figure~\ref{fig:4frame} shows the visibility amplitude and phase images produced by the Fourier transform of the interferogram, for the $3''$ mirror rotation data. Each visibility sample is labeled by the corresponding baseline. The visibility amplitudes and phases for each sample were derived using a vector sum over the area of the sample out to 30\% intensity (4 pixel radius in the $u$--$v$ coordinate system).

The self-calibration process was then employed, as delineated in Section~\ref{sec:method}, for each frame in a 30-exposure experiment. In this case, we have a very good understanding of the size and shape of the synchrotron light source (Gaussian\cite{Nikolic2024, iriso2024}), and hence a single self-calibration iteration was employed to derive the voltage complex gains. Note that phase wraps are reconciled by considering continuity of the planar wavefront solution across the mask. The visibility data were analyzed using the CASA data reduction software package \cite{casa:2017}. CASA has extensive tools for interferometric data self-calibration and analysis, long used in radio astronomy and in other areas. 

\begin{figure}[!htb]
\centering 
\centerline{\includegraphics[scale=0.24]{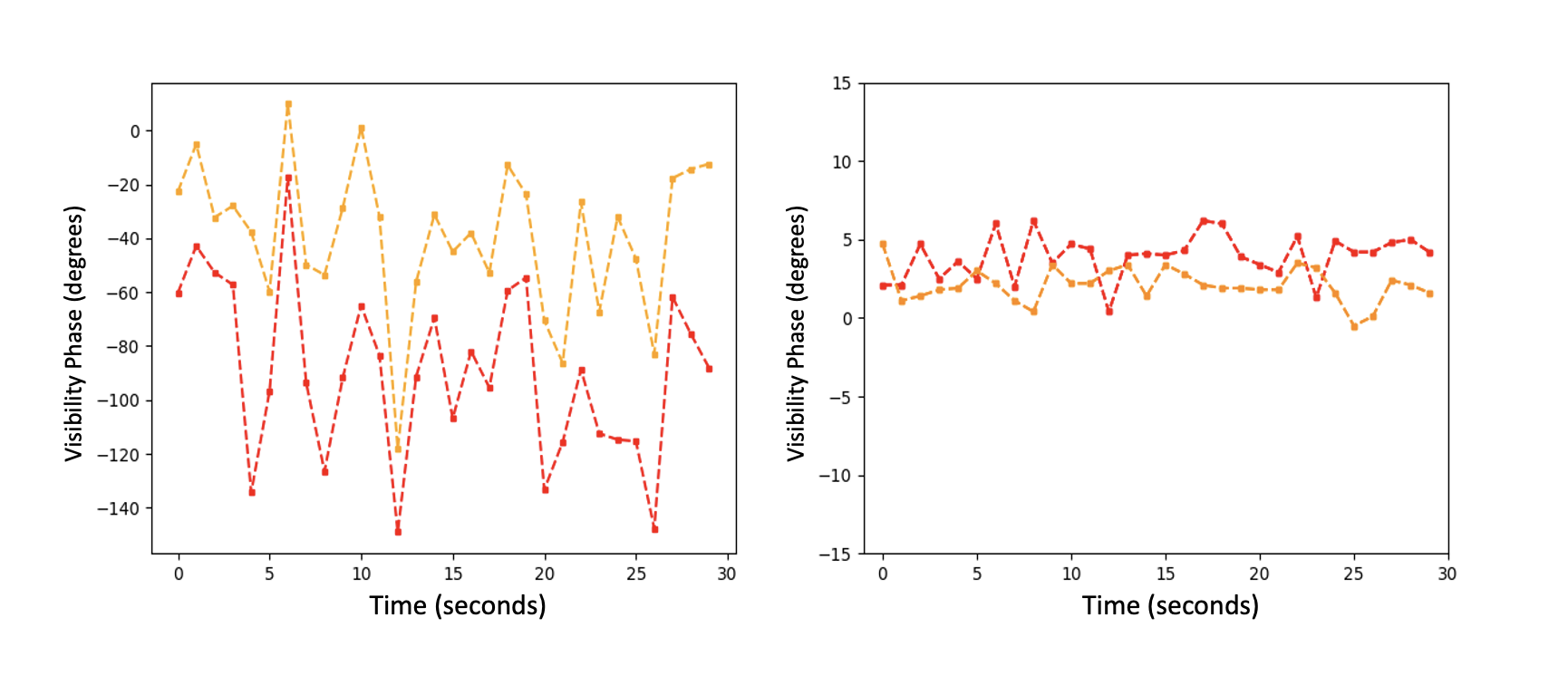}}
\caption{The visibility phase time series of baselines 1-6 (red) and 2-5 (orange). These are two neighboring and parallel baselines that do not have a hole in common (see Figure~\ref{fig:4frame}). Left shows the phases for the u,v points before self-calibration. Note that the Y-axis range is much larger for the left-hand figure.  
}
\label{fig:25.16}
\end{figure}

Figure~\ref{fig:25.16} shows a comparison of the visibility phases on two neighboring, roughly parallel baselines, namely 1-6 and 2-5, both before and after calibration. Before calibration, the mean values for the visibility phases are systematically offset from zero phase. Further, the phase fluctuations of the time series are large (rms $= 30^\circ$), and closely correlated in time between the two baselines, even though the baselines do not share a hole. After calibration, the rms variation in Figure~\ref{fig:25.16} over the 30 time samples $= 1.2^\circ$, and there is no residual correlation between baselines. 

The analysis of Figure~\ref{fig:25.16} has two important implications. First, the fact that before calibration the visibility phase fluctuations correlate between close, parallel baselines that do not share a hole implies that the pre-calibration phase jitter is not due to a random noise term, such as photon counting statistics, but reflects some systematics of propagation that has spatial correlation, such as vibration of optical components or lab turbulence. And second, the fact that the noise-like post-calibration phase jitter is a factor 30 lower than pre-calibration phase jitter, implies that the vibrational or turbulence induced phase jitter can be factored into element-based corruptions, that these corruptions are being measured at high signal-to-noise. 

\begin{figure}[!htb]
\centering 
\centerline{\includegraphics[scale=0.6]{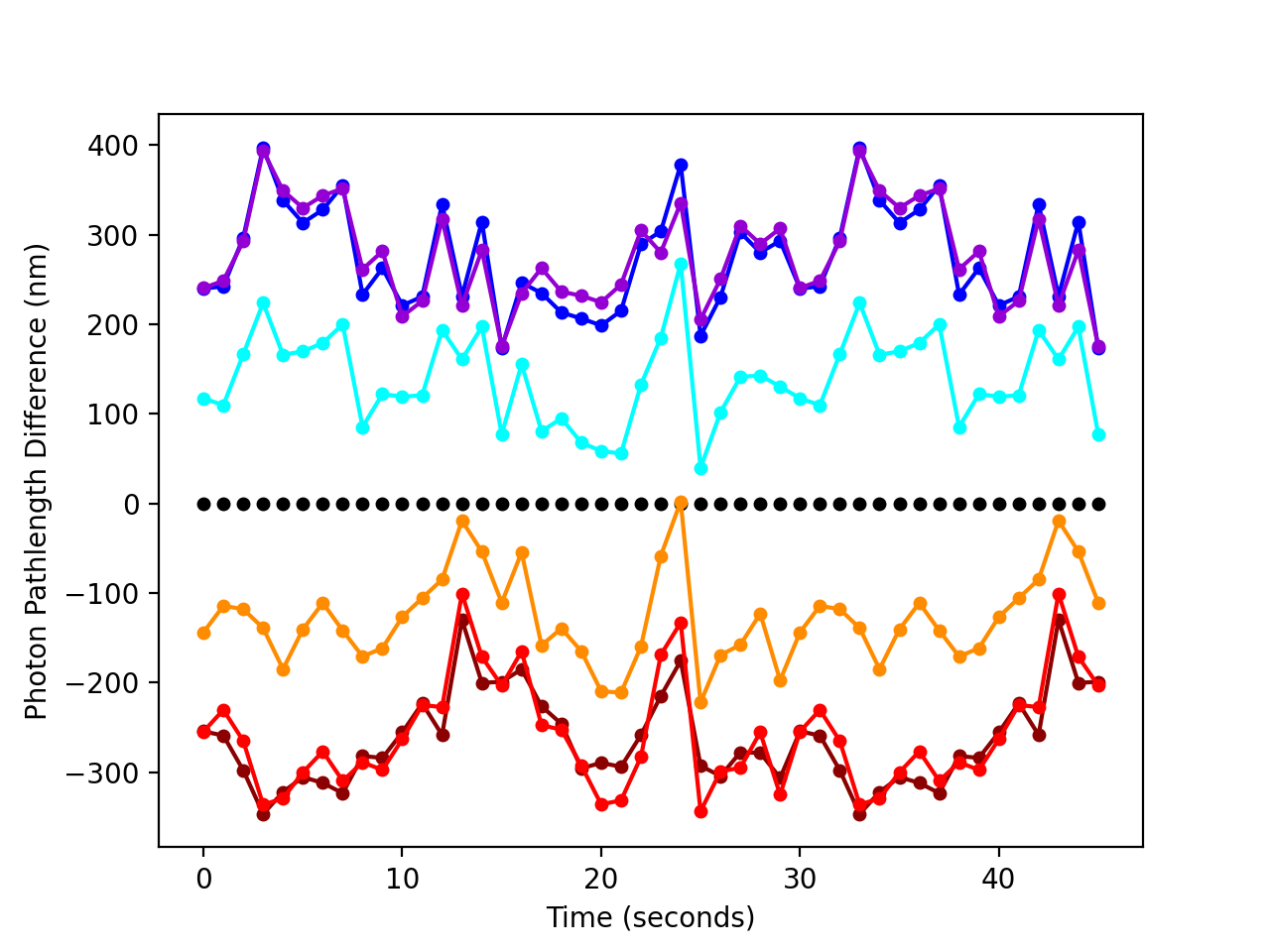}}
\caption{Time series of 30 measurements of 1~ms duration spaced by 1~s for photon path-lengths through the optical system for the rotation $3''$ dataset, as derived from the self-calibration phase gains using Equ.~\ref{eqn:path}. The black points are for hole 0, which is the reference position in the self-calibration, and hence zero by construction. Holes 1,2,3,4,5,6 correspond to colors dark red, red, orange, cyan, blue, violet, respectively. The path-length jitter with time has a typical rms between $\sim 30$~nm to 50~nm. The mean path-length values to each hole reflect the rotated mirror setting.
}
\vspace*{0.5cm}
\label{fig:3as.path}
\end{figure}

Figure~\ref{fig:3as.path} shows the derived path-length time series calculated from the gain phases using Equ.~\ref{eqn:path}, for the $3''$ rotation data. There is substantial `jitter' in the path-lengths, or gain phases, with an rms between $\sim 30$~nm to 50~nm. Again, this jitter is due to vibration of the optics, with a possible contribution from laboratory turbulence. The path-length jitter is strongly correlated between neighboring holes (e.g. 5 and 6), and less so for holes on opposite sides of the mask (e.g. 1 and 6). A cross-correlation analysis of the time series shows that gain phases for closely separated holes are well correlated at $> 5\sigma$ for the zero-lag point, while more widely separated holes show no clear correlation in phase gains, $< 2\sigma$ at zero lag, where $\sigma$ is derived as the rms scatter of the non-zero lag points.

The mean values for the path-lengths are clearly non-zero. These mean values reflect the tilt of the mirror around the Y-axis, such that hole 5 and 6, which are at the same X-distance from hole 0 (the reference hole for self-calibration), show the same tilt-induced path-lengths, and likewise for holes 1 and 2, but now of opposite sign since these hole are on the other side of the mask from 5 and 6. Holes 3 and 4, at intermediate X distances, fall in-between in the path-length derivation. We analyze the rotating mirror-induced wavefront tilt across the mask by fitting planes in 3-dimensions to the path-length measurements to get the X (horizontal) and Y (vertical) tilts of the wavefront. The tilt results will be considered in Section~\ref{sec:tilt}. We first consider the static, non-planar metrology, which must first be removed to optimize the planar tilt analysis.

\section{Static non-planar metrology}
\label{sec:static}

We consider the static, non-planar metrology of the system, meaning path-length delays that are constant with time but not well-fit by just a tip-tilt plane. First, in terms of the static tip-tilt term, again we have found that the Airy disk centering pixel position is constant for the three $0''$ rotation settings to within the setting accuracy of the device of $\sim 0.3''$ (Section~\ref{sec:processing}).

\begin{figure}[!htb]
\centering 
\centerline{\includegraphics[scale=0.24]{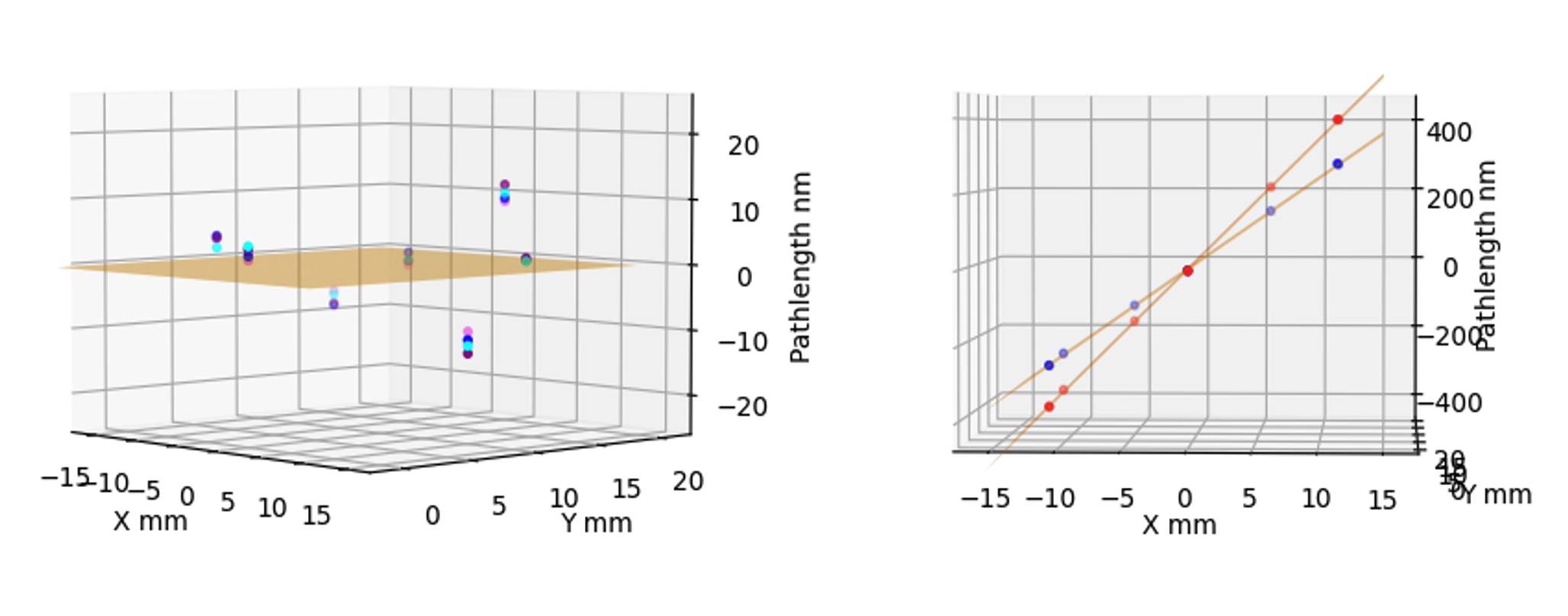}}
\caption{{\bf Left}: 3D projections of the path-lengths to each of the 7 holes derived for the three $0''$ rotation angle data sets (blue, cyan, purple), after subtracting the best fit plane in each case. Also shown are points for the  $2''$ mirror rotation data, after subtracting the best fit plane (violet). The displayed plane in orange is the zero tilt plane for reference. These offsets correspond to the static metrology terms of the photon path-lengths through the system that cannot be fit by a planar term, but would require higher order wavefront curvature terms.  The small scatter in the 4 measurements at each hole implies that the static metrology terms repeat to within $\pm 1~$nm for all the data sets. 
{\bf Right}: 3D projections of the data and best fit plane for the $3''$ and $4''$ mirror rotations, shown in an edge on projection to highlight how well the points fit a plane after removal of the static metrology derived from the $0''$ rotation data. The rms residuals relative to the best fit plane are $\sim 2$~nm for all four rotation angle data. Note the tilt angle is greatly exaggerated in the figure, since the path-length scale (Z) is in nanometers, while the X and Y mask scale is in millimeters. For example, a $4''$ wavefront tilt $= 1.9\times 10^{-5}$~radians. 
}
\vspace*{0.5cm}
\label{fig:Met.34Edge}
\end{figure}

However, there are also non-planar terms, meaning higher order optics curvature beyond tip-tilt. We have investigated these by looking at the residuals of the path-lengths to each hole after removing the dominant static tip-tilt term. Table~\ref{table:metrology} summarizes the results, and the results are plotted in Figure~\ref{fig:Met.34Edge}. We include data from the series of three experiments with $0''$ rotation (two at the start and one at the end), as well as the $2''$ rotation experimental data after removing the best fit tip-tilt due to the rotated mirror. 

In all cases, we find that the non-tip-tilt metrology terms for the holes have mean values for the time series that range from $\sim 2~$nm up to $12$~nm, and these values repeat to better than to $\pm 1$~nm for the four experiments. Figure~\ref{fig:Met.34Edge} shows that the scatter in the 4 measurements is typically much smaller than the departure of the mean measurement from the plane, implying static non-planar distortions in the optics at the level of up to $\sim 12$~nm. 

\begin{table}
\centering
\footnotesize
\caption{Static Non-Planar Metrology}
\begin{tabular}{lccccccc} 
  \hline
  \hline
  Time & Ap 0 & Ap 1 & Ap 2  & Ap 3 & Ap 4 & Ap 5  & Ap 6 \\
~  & nm & nm & nm & nm & nm & nm & nm \\
\hline
$0''$ 1st Series & 2.45 & 4.01 & -8.11 & 0.52 & 9.95 & 1.27 & -10.10 \\
$0''$ 2nd Series & 3.92 & 2.18 & -6.40 & -0.57 & 10.78 & 1.06 & -10.96 \\
$0''$ End Series & 3.49 & 3.65 & -7.78 & -0.82 & 12.04 & 1.53 & -12.10 \\
$2''$ & 1.85 & 3.65 & -5.83 & -1.47 & 9.45 & 1.19 & -8.83 \\
\hline
\hline
\vspace{0.1cm}
\end{tabular}
\label{table:metrology}
\end{table}

These non-planar static terms correspond to higher order curvature in the wavefront beyond tip-tilt. We have attempted fitting for this curvature with Zernike polynomials\cite{vacalebre2022}, however, our current aperture mask has only 7 holes, and hence the results for modes beyond tip-tilt are poorly constrained and show considerable covariance. Further tests are planned with a mask with more holes.

In the following analysis of rotated mirror data (Section~\ref{sec:tilt}), the mean static metrology terms derived from the unrotated ($0''$) data, are first subtracted, in order to isolate the the tip-tilt derivation. Note that these static, non-planar terms are much smaller than the path-lengths induced by rotating the mirror, by typically an order of magnitude or more (see Figure~\ref{fig:4frame}).

\section{Tip-Tilt Analysis}
\label{sec:tilt}

\subsection{Rotating mirror tests}

\begin{figure}[!htb]
\centering 
\centerline{\includegraphics[scale=0.25]{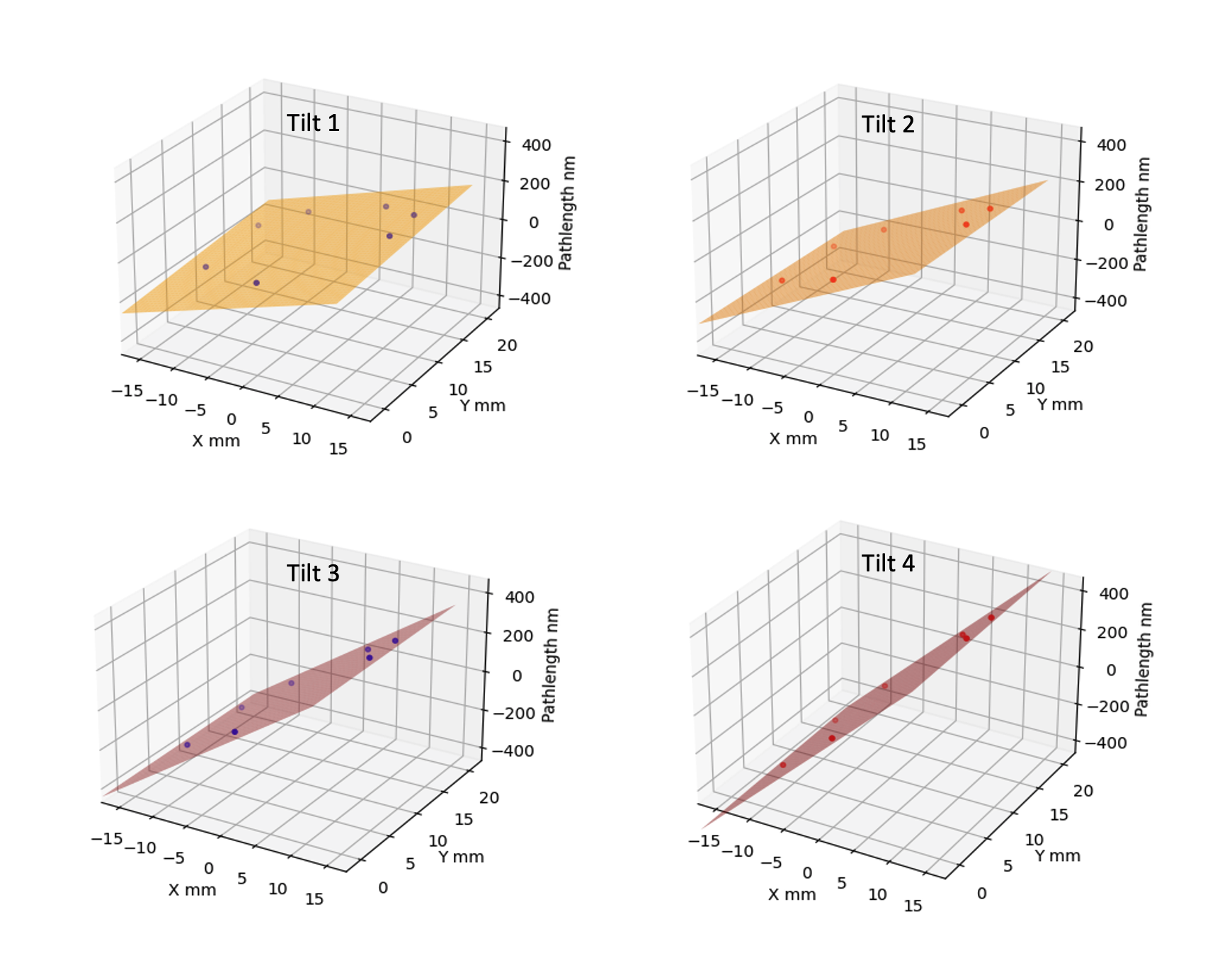}}
\caption{3D projections of the data (7 points in each frame) and best fit plane for the four mirror rotations. The scales are the same for all four frames. Again, the tilt angle is greatly exaggerated in the figure, since the path-length scale is in nm while the X and Y mask scale is in mm. 
}
\vspace*{0.5cm}
\label{fig:tilt1-4}
\end{figure}

We next consider the rotated mirror datasets. Figure~\ref{fig:tilt1-4} shows 3D projections of the fit planes from the rotated mirror data, and values for the best fit plane tilt are listed in Table~\ref{table:WFTilt}. Again, the accuracy of mirror rotation angle setting is, at best, $\sim 0.3''$ (Section~\ref{sec:experiment}). However, we have three independent methods with which to derive the wavefront tilt, as follows: 

The first method is in the image plane using the centering pixel position of the Airy disk. The image shift in arcseconds is equal to the wavefront tilt. Hence, we compare mean centering pixel positions from the 30 exposure time series of the rotated mirror data with the values derived from the unrotated mirror data. 

The second method involves using phase gradients across u,v samples, or visibilities, after frame centering using the pixel center derived from the unrotated mirror data. Inspection of the visibility phase image in Figure~\ref{fig:4frame} shows phase gradients across each u,v sample. These phase gradients are of similar magnitude and orientation for each sample, and they reflect the tilt of the wavefront, as was shown in Cheetham et al.\cite{Cheetham:12} (they use the term `splodge' for the visibility samples in the $u$--$v$ coordinates). Qualitatively, the phase gradients across these splodges can be considered sub-structure relating to the finite size of the holes, such that a value at one edge of a visibility splodge could be considered the correlation between photons on the near edge of one hole with the far edge of the second hole.

We have fit planes to the phase gradient across the u,v samples, in which the slopes are derived in units of radians per u,v cell. Hence, a one pixel shift in the image implies a slope across the u,v sample of $2\pi/2048 = 0.00307$. For example, if our slope fitting finds a value of 0.04 radians per u,v pixel, then the shift in the image plane $= .04/.00307 = 13$ image pixels. Using a pixel size of $0.153''$  then leads to a wavefront tilt $= 13 \times 0.153'' = 2''$. We have considered using all the u,v samples to derive the mean wavefront slope, or just using the 0,0 sample, corresponding to the total power. The latter has roughly ten times the number of photons than the interferometric baseline samples. We find the results are similar, but the scatter in the time series using just the 0,0 sample is smaller, and we list those results in Table~\ref{table:WFTilt}. 

The third method is the self-calibration phase gain solution method described in Section~\ref{sec:method}, where the wavefront tip-tilt is derived by planar fitting to the photon path-lengths to each hole. 

Table~\ref{table:WFTilt} shows the results for all three methods of the mean wavefront tilts in X and Y.  The error-bars indicate the rms scatter in the 30 measurements of the time series divided by $\sqrt{30}$. In all cases error-bars are $\sim 0.1''$. 

We find that the X tilts agree to within $\sim 0.1''$ for all three methods. This result lends confidence that the measurements are reliable, and further, that the self-calibration method recovers the wavefront tilt to this level of accuracy. We find that the Y tilts are small, mostly consistent with zero.  
Lastly, Figure~\ref{fig:Met.34Edge} also shows the 3D projection for the $3''$ and $4''$ mirror rotation data but now in an edge-on view. This figure is a graphic illustration that, after subtraction of the static metrology path-length derived from the $0''$ rotation data, the points fit very closely to a plane, with an rms scatter around the best fit plane of $\sim 2$~nm for all of the rotated mirror data. 

\begin{table}
\centering
\footnotesize
\caption{Wavefront Tilt}
\begin{tabular}{lcccccc} 
  \hline
  \hline
  Mirror & Selfcal & Image Centering  & u,v sample slope & Selfcal & Image Centering  & u,v sample slope \\
rotation & X arcsec & X arcsec & X arcsec & Y arcsec & Y arcsec & Y arcsec \\
\hline
$1''$ & $2.15\pm 0.09$ & $2.30\pm 0.13$ & $2.15 \pm 0.09$ & $0.56 \pm 0.09$ & $0.66\pm 0.13$ & $0.70 \pm 0.09$ \\
$2''$ & $3.27 \pm 0.12$ & $3.21\pm 0.14$ & $3.17\pm 0.12$ & $-0.07\pm 0.13$ & $-0.01\pm 0.14$ & $0.06 \pm 0.12$ \\
$3''$ & $4.88\pm 0.1$ & $4.77\pm 0.13$ & $4.78\pm 0.12$ & $0.08\pm 0.1$ & $-0.04\pm 0.13$ & $0.11\pm 0.11$ \\
$4''$ & $6.96\pm 0.1$ & $6.92\pm 0.10$ & $6.95\pm 0.10$ & $0.24 \pm 0.1$ & $0.27\pm 0.10$ & $0.39\pm 0.10$ \\
\hline
\hline
\vspace{0.1cm}
\end{tabular}
\label{table:WFTilt}
\end{table}

\subsection{Temporal variations: adaptive optics}

\begin{figure}[!htb]
\centering 
\centerline{\includegraphics[scale=0.3]{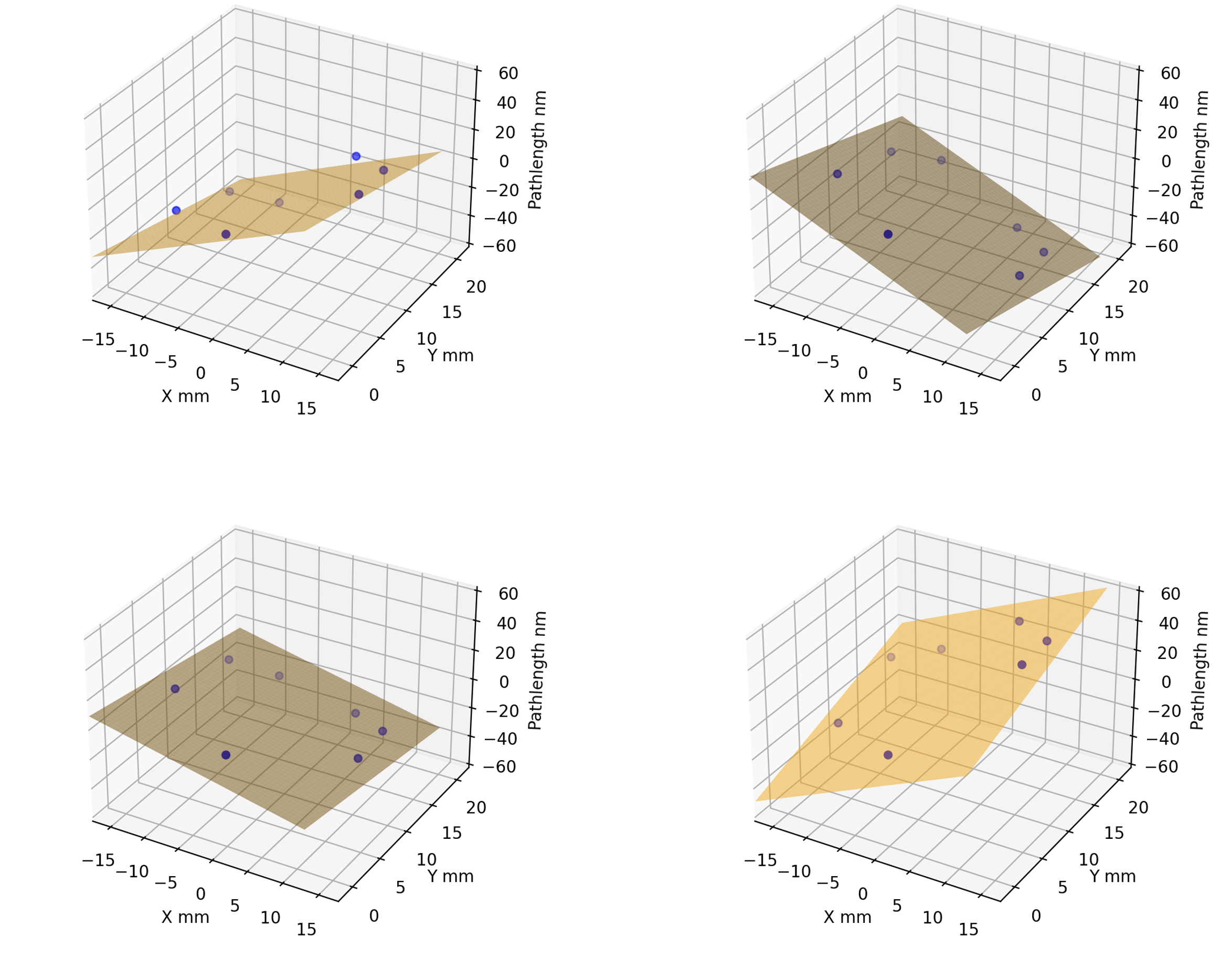}}
\caption{Three-dimensional projections of the residual photon path-length variations for four frames from the 7-hole mask time series of measurements for $0''$ mirror rotation data. These path-length residuals are calculated from self-calibration gain phases after subtracting the mean gain phases derived from the full 30 frame time series, and then converted to path-length in nanometers using Equation~\ref{eqn:path}. These residual photon path-lengths represent the jitter of the wavefront from frame to frame (see Figure:\ref{fig:3as.path}). Also shown is a best-fit planar wavefront to the measurements, corresponding to the variation of the tip-tilt term for the wavefront with time. 
}
\label{fig:TILT}
\end{figure}

Figure~\ref{fig:TILT} shows the result of a planar tip-tilt fit to the self-calibration derived path-lengths for the 7-hole mask for four individual 1~ms exposures in one of the $0''$ rotation experiments. The planar fits show path-length excesses of up to $\sim 50$~nm across the mask. A 50~nm wavefront tilt across a 22~mm diameter mask implies a wavefront tilt of $0.5''$, or $\sim 2.3\times 10^{-6}$~radians. The fitted tilt plane orientation varies randomly between frames taken every one second, as expected since the jitter occurs on millisecond timescales, as implied by analysis of the visibility coherence with exposure time \cite{Carilli2024}. 

\subsection{Noise, Precision, Dynamic Range, and Sampling}
\label{sec:precision}

In our aperture mask wavefront sensing approach, the path-length precision per frame is set by the precision with which the phase gain can be determined by self-calibration. This precision, in turn, depends on the phase noise per visibility. 

In our experiment the photon counts per visibility are a few $\times 10^6$, so the $\sqrt{N}$ noise due to photon counting statistics per visibility should be of order $5\times 10^{-4}$, which, in phase implies $\sim 0.2^\circ$. Empirically, our analysis shows an rms for the 30-frame time series of the visibility phases after calibration of $1.2^\circ$ (see Figure~\ref{fig:25.16}). 

There are steps in the processing that might add a random noise-like term, other than the minimum set by photon counting statistics\cite{Nikolic2024}. One possible contribution comes from the pixelization of the image: the phase of a visibility is the shift of the fringe pattern on the CCD relative to the reference pixel, and the accuracy with which this can be measured is determined by the pixel size and signal to noise. Others include truncating of the interferogram by the finite size of the CCD, and the adopted radius in the u,v plane for deriving complex visibilities.

Our empirical estimate of the visibility-based phase noise of $1.2^\circ$ then sets the noise-floor for the gain phases derived by the self-calibration process, as follows: In our 7-hole experiment, each gain phase is derived from 6 visibilities. Assuming Gaussian random errors, we expect the noise on the gain phases to decrease relative to the visibility phase noise by a factor $\sqrt{6}$. Hence, the noise-like limit to the gain phase measurements per frame is $\sim 0.5^\circ$. At 400~nm, this gain phase rms limit then  implies a path-length precision of 0.6~nm.

Turning to dynamic range, defined as the maximum wavefront distortion measurable across the aperture, in our aperture mask approach the maximum phase change between holes is limited by a phase wrap of $360^\circ$, or one wavelength between pairs of holes. Interpolating solutions between holes across the mask can reconcile phase wraps, but that will only increase the dynamic range to a few wavelengths, depending on the hole size and aperture sampling. 

The dynamic range can be greatly enhanced by adding a frequency axis to the analysis. Gain phases for a physical path-length distortion are linear in wavelength, so making a closely spaced measurement in frequency can determine the phase wraps. For example, making two measurements of the phase screen separated in wavelength by 5\% would increase the dynamic range by a factor 20 relative to a single measurement.  Using the frequency axis to determine phase wraps is a well known technique in radio Very Long Baseline Interferometry, where interference fringes are generated between antennas separated by up to 10,000~km. VLBI fringe fitting determines path-length delays directly by using visibility spectral measurements with fine frequency channel resolution. Linear fits to the phase slope with frequency then determine the path-length delay \cite{walker1999, schwabcotton1983}. 

Lastly, the sampling of the aperture plane of a 7-hole mask is relatively sparse, thereby precluding derivation of higher order modes for the wavefront beyond tip-tilt (Section~\ref{sec:static}). Manufacturing masks with more holes is trivial, however, maintaining non-redundancy becomes an interesting mathematical problem\cite{gonzalez2011, mckay2022}. Of course, some non-redundancy can be handled in the self-calibration process by simply omitting the redundant samples from the fitting, given the process is typically highly over-constrained, or altering the self-calibration equations to allow for redundant samples.

\section{Summary}
\label{sec:summary}

We have demonstrated the efficacy of a new method of wavefront sensing using optical masking interferometry with self-calibration to derive photon path-length delays through the optical system. Tests were performed at the ALBA Synchrotron light source optical laboratory.  For this initial quantitative demonstration, our analysis focused on tip-tilt distortions using a rotating mirror, for which we have two other independent methods to calibrate the wavefront tilt results. A comparison of the self-calibration wavefront sensing with the two independent methods results in agreement of all methods to within $0.1''$, or $4.9\times 10^{-7}$~radians, in deriving the wavefront tilt induced by the rotating mirror.  

We also determine the static metrology for the optical system. The static non-planar wavefront path-length distortions range up to 12~nm at a given hole, and are repeatable to within $\pm 1$~nm  for all the datasets. 

We emphasize that this new wavefront sensing method measures the photon path-lengths to each hole in the mask relative to the path-length to a reference position on the mask. The method is not a gradient, or angle-based measurement\cite{neal2002,Mansuripur2009,yi2021}, nor a reference wave method\cite{sheldovka,goodwinwyant}, nor a generalized Foucault knife-edge test\cite{shatokhina}. The method does not require a reference wave or surface, has no moving parts, and is agnostic to source shape since the self-calibration process can derive source shape and wavefront shape in a joint optimization. We have demonstrated experimentally that wavefront distortions can be measured at the $\sim 1$~nm level, with the possibility of going a factor six lower if the photon counting noise can be reached. The data processing per frame can be done in milliseconds\cite{Nikolic2024}. 

\acknowledgments 
The National Radio Astronomy Observatory is a facility of the National Science Foundation operated under cooperative agreement by Associated Universities, Inc.. Image processing was performed using the Software: Astronomical Image Processing System (AIPS) \cite{Greisen2003} and Common Astronomical Software Applications (CASA) \cite{casa:2017}. 
Related patents and patent applications: Patent No. US 12,104,901 B2; Provisional Patent No. 63/355,174 (RL 8127.032.USPR); Provisional Patent No. 63/648,303 (RL 8127.306.USPR).

\bibliography{report} 

\begin{thebibliography}{10}

\bibitem{geary1995}
Geary, J.~M.,  [{\em Introduction to Wavefront Sensors}{\nolinebreak\hspace{0.1em}]}, vol.~18 of {\em Tutorial Texts in Optical Engineering}, SPIE Press (1995).

\bibitem{Watnik18}
Watnik, A.~T. and Gardner, D.~F., ``Wavefront sensing in deep turbulence,'' {\em Opt. Photon. News}~{\bf 29},  38--45 (Oct 2018).

\bibitem{holmes2022}
Holmes, R.~B., ``Adaptive optics for directed energy: Fundamentals and methodology,'' {\em AIAA Journal}~{\bf 60}(10),  5633--5644 (2022).

\bibitem{forest2004}
{Forest}, C.~R., {Canizares}, C.~R., {Neal}, D.~R., {McGuirk}, M., and {Schattenburg}, M.~L., ``{Metrology of thin transparent optics using Shack-Hartmann wavefront sensing},'' {\em Optical Engineering}~{\bf 43},  742--753 (Mar. 2004).

\bibitem{rammage2002}
Rammage, R.~R., Neal, D.~R., and Copland, R.~J., ``{Application of Shack-Hartmann wavefront sensing technology to transmissive optic metrology},'' in [{\em Advanced Characterization Techniques for Optical, Semiconductor, and Data Storage Components}{\nolinebreak\hspace{0.1em}]},  Duparr{\'e}, A. and Singh, B., eds.,  {\bf 4779},  161 -- 172, International Society for Optics and Photonics, SPIE (2002).

\bibitem{adapa2020}
Adapa, B.~R., {\em {The application of wavefront sensing methods to optical surface metrology}}, theses, {Universit{\'e} Grenoble Alpes []} (Oct. 2020).

\bibitem{Liang1994}
Liang, J., Grimm, B., Goelz, S., and Bille, J.~F., ``Objective measurement of wave aberrations of the human eye with the use of a hartmann--shack wave-front sensor,'' {\em J. Opt. Soc. Am. A}~{\bf 11},  1949--1957 (Jul 1994).

\bibitem{neal2002}
Neal, D.~R., Copland, J., and Neal, D.~A., ``{Shack-Hartmann wavefront sensor precision and accuracy},'' in [{\em Advanced Characterization Techniques for Optical, Semiconductor, and Data Storage Components}{\nolinebreak\hspace{0.1em}]},  Duparr{\'e}, A. and Singh, B., eds.,  {\bf 4779},  148 -- 160, International Society for Optics and Photonics, SPIE (2002).

\bibitem{Readhead+1988}
{Readhead}, A.~C.~S., {Nakajima}, T.~S., {Pearson}, T.~J., {Neugebauer}, G., {Oke}, J.~B., and {Sargent}, W.~L.~W., ``{Diffraction-Limited Imaging with Ground-Based Optical Telescopes},'' {\em AJ}~{\bf 95},  1278 (Apr. 1988).

\bibitem{Schwab1980}
{Schwab}, F.~R., ``{Processing of three-dimensional data},'' in [{\em 1980 International Optical Computing Conference I}{\nolinebreak\hspace{0.1em}]},  {Rhodes}, W.~T., ed., {\em Society of Photo-Optical Instrumentation Engineers (SPIE) Conference Series} {\bf 231},  18 (Jan. 1980).

\bibitem{Schwab1981}
{Schwab}, F.~R., ``{Robust Solution for Antenna Gains},'' {\em NRAO VLA Scientific Memoranda}~{\bf 136} (1981).

\bibitem{Readhead+Wilkinson1978}
{Readhead}, A.~C.~S. and {Wilkinson}, P.~N., ``{The mapping of compact radio sources from VLBI data.},'' {\em ApJ}~{\bf 223},  25--36 (July 1978).

\bibitem{Cornwell+Wilkinson1981}
{Cornwell}, T.~J. and {Wilkinson}, P.~N., ``{A new method for making maps with unstable radio interferometers},'' {\em MNRAS}~{\bf 196},  1067--1086 (Sept. 1981).

\bibitem{Nikolic2024}
{Nikolic}, B., {Carilli}, C.~L., {Thyagarajan}, N., {Torino}, L., and {Iriso}, U., ``{Two-dimensional synchrotron beam characterization from a single interferogram},'' {\em Physical Review Accelerators and Beams}~{\bf 27},  112802 (Nov. 2024).

\bibitem{iriso2024}
{Iriso}, U., {Torino}, L., {Carilli}, C., {Nikolic}, B., and {Thyagarajan}, N., ``{New interferometric aperture masking technique for full transverse beam characterization using synchrotron radiation},'' {\em arXiv e-prints} ,  arXiv:2409.11135 (Sept. 2024).

\bibitem{Thyagarajan+2025}
{Thyagarajan}, N., {Nikolic}, B., {Carilli}, C.~L., , {Torino}, L., and {Iriso}, U., ``{Two-dimensional Synchrotron Beam Shape Characterization using Interferometric Closure Amplitudes},'' {\em in-prep}  (2025).

\bibitem{Cornwell+Fomalont1999}
{Cornwell}, T. and {Fomalont}, E.~B., ``{Self-Calibration},'' in [{\em Synthesis Imaging in Radio Astronomy II}{\nolinebreak\hspace{0.1em}]},  {Taylor}, G.~B., {Carilli}, C.~L., and {Perley}, R.~A., eds., {\em Astronomical Society of the Pacific Conference Series} {\bf 180},  187 (Jan. 1999).

\bibitem{Pearson+Readhead1984}
{Pearson}, T.~J. and {Readhead}, A.~C.~S., ``{Image Formation by Self-Calibration in Radio Astronomy},'' {\em Annual Reviews of Astronomy and Astrophysics}~{\bf 22},  97--130 (Jan. 1984).

\bibitem{Perley1999}
{Perley}, R.~A., ``{High Dynamic Range Imaging},'' in [{\em Synthesis Imaging in Radio Astronomy II}{\nolinebreak\hspace{0.1em}]},  {Taylor}, G.~B., {Carilli}, C.~L., and {Perley}, R.~A., eds., {\em Astronomical Society of the Pacific Conference Series} {\bf 180},  275 (Jan. 1999).

\bibitem{TMS2017}
{Thompson}, A.~R., {Moran}, J.~M., and {Swenson}, George~W., J.,  [{\em {Interferometry and Synthesis in Radio Astronomy, 3rd Edition}}{\nolinebreak\hspace{0.1em}]}, Springer, Cham (2017).

\bibitem{perley1984}
{Perley}, R.~A., {Dreher}, J.~W., and {Cowan}, J.~J., ``{The jet and filaments in Cygnus A.},'' {\em Astrophysical Journal Letters}~{\bf 285},  L35--L38 (Oct. 1984).

\bibitem{davies2012}
{Davies}, R. and {Kasper}, M., ``{Adaptive Optics for Astronomy},'' {\em Annual Reviews of Astronomy and Astrophysics}~{\bf 50},  305--351 (Sept. 2012).

\bibitem{sheldovka}
Sheldakova, J., Kudryashov, A., Zavalova, V., and Romanov, P., ``{Shack-Hartmann wavefront sensor versus Fizeau interferometer for laser beam measurements},'' in [{\em Laser Resonators and Beam Control XI}{\nolinebreak\hspace{0.1em}]},  Kudryashov, A.~V., Paxton, A.~H., Ilchenko, V.~S., and Aschke, L., eds.,  {\bf 7194},  71940B, International Society for Optics and Photonics, SPIE (2009).

\bibitem{goodwinwyant}
{Goodwin}, E.~P. and {Wyant}, J.~C.,  [{\em {Field guide to interferometric optical testing}}{\nolinebreak\hspace{0.1em}]} (2006).

\bibitem{Mansuripur2009}
Mansuripur, M.,  [{\em The Shack–Hartmann wavefront sensor}{\nolinebreak\hspace{0.1em}]},  624–631, Cambridge University Press (2009).

\bibitem{yi2021}
{Yi}, S., {Xiang}, J., {Zhou}, M., {Wu}, Z., {Yang}, L., and {Yu}, Z., ``{Angle-based wavefront sensing enabled by the near fields of flat optics},'' {\em Nature Communications}~{\bf 12},  6002 (Oct. 2021).

\bibitem{shatokhina}
Shatokhina, I., Hutterer, V., and Ramlau, R., ``{Review on methods for wavefront reconstruction from pyramid wavefront sensor data},'' {\em Journal of Astronomical Telescopes, Instruments, and Systems}~{\bf 6}(1),  010901 (2020).

\bibitem{vacalebre2022}
Vacalebre, M., Frison, R., Corsaro, C., Neri, F., Conoci, S., Anastasi, E., Curatolo, M.~C., and Fazio, E., ``Advanced optical wavefront technologies to improve patient quality of vision and meet clinical requests,'' {\em Polymers}~{\bf 14}(23) (2022).

\bibitem{Lombardo2009}
Lombardo, M. and Lombardo, G., ``New methods and techniques for sensing the wave aberrations of human eyes,'' {\em Clinical and Experimental Optometry}~{\bf 92}(3),  176--186 (2009).

\bibitem{vanCittert34}
{van Cittert}, P.~H., ``{Die Wahrscheinliche Schwingungsverteilung in Einer von Einer Lichtquelle Direkt Oder Mittels Einer Linse Beleuchteten Ebene},'' {\em Physica}~{\bf 1},  201--210 (1934).

\bibitem{Zernike38}
{Zernike}, F., ``{The concept of degree of coherence and its application to optical problems},'' {\em Physica}~{\bf 5},  785--795 (Aug. 1938).

\bibitem{Born+Wolf1999}
{Born}, M. and {Wolf}, E.,  [{\em {Principles of Optics}}{\nolinebreak\hspace{0.1em}]} (1999).

\bibitem{young1803}
{Young}, T.,  [{\em {The bakerian lecture : Experiments and calculations relative to physical optics}}{\nolinebreak\hspace{0.1em}]} (1803).

\bibitem{JWST}
{Sivaramakrishnan}, A., {Tuthill}, P., {Lloyd}, J.~P., {Greenbaum}, A.~Z., {Thatte}, D., {Cooper}, R.~A., and {Vandal}, e.~a., ``{The Near Infrared Imager and Slitless Spectrograph for the James Webb Space Telescope. IV. Aperture Masking Interferometry},'' {\em Publication of the Astronomical Society of the Pacific}~{\bf 135},  015003 (Jan. 2023).

\bibitem{Torino2016}
Torino, L. and Iriso, U., ``Transverse beam profile reconstruction using synchrotron radiation interferometry,'' {\em Phys. Rev. Accel. Beams}~{\bf 19},  122801 (Dec 2016).

\bibitem{Carilli2024}
Carilli, C., Nikolic, B., Torino, L., Iriso, U., and Thyagarajan, N., ``{Deriving the size and shape of the ALBA synchrotron light source with optical aperture masking: technical choices},'' tech. rep., ALBA, arXiv:2406.02114 (2 2024).

\bibitem{casa:2017}
{McMullin}, J.~P., {Waters}, B., {Schiebel}, D., {Young}, W., and {Golap}, K., ``{CASA Architecture and Applications},'' in [{\em Astronomical Data Analysis Software and Systems XVI}{\nolinebreak\hspace{0.1em}]},  {Shaw}, R.~A., {Hill}, F., and {Bell}, D.~J., eds., {\em Astronomical Society of the Pacific Conference Series} {\bf 376},  127 (Oct. 2007).

\bibitem{Cheetham:12}
Cheetham, A.~C., Tuthill, P.~G., Sivaramakrishnan, A., and Lloyd, J.~P., ``Fizeau interferometric cophasing of segmented mirrors,'' {\em Opt. Express}~{\bf 20},  29457--29471 (Dec 2012).

\bibitem{walker1999}
{Walker}, R.~C., ``{Very Long Baseline Interferometry},'' in [{\em Synthesis Imaging in Radio Astronomy II}{\nolinebreak\hspace{0.1em}]},  {Taylor}, G.~B., {Carilli}, C.~L., and {Perley}, R.~A., eds., {\em Astronomical Society of the Pacific Conference Series} {\bf 180},  433 (Jan. 1999).

\bibitem{schwabcotton1983}
Schwab, F. and Cotton, W., ``Global fringe search techniques for vlbi,'' {\em The Astronomical Journal}~{\bf 88},  688--694 (04 1983).

\bibitem{gonzalez2011}
Gonz\'{a}lez, A.~I. and Mej\'{i}a, Y., ``Nonredundant array of apertures to measure the spatial coherence in two dimensions with only one interferogram,'' {\em J. Opt. Soc. Am. A}~{\bf 28},  1107--1113 (Jun 2011).

\bibitem{mckay2022}
McKay, D., Grydeland, T., and Gustavsson, B., ``Manx arrays: Perfect non-redundant interferometric geometries,'' {\em Radio Science}~{\bf 57}(9),  e2022RS007500 (2022).
\newblock e2022RS007500 2022RS007500.

\bibitem{Greisen2003}
{Greisen}, E.~W.,  [{\em {AIPS, the VLA, and the VLBA}}{\nolinebreak\hspace{0.1em}]}, vol.~285,  109 (2003).

\end{thebibliography}
\bibliographystyle{spiebib} 

\end{document}